\documentclass[a4paper,landscape,twocolumn,9pt]{article}

\usepackage{extsizes}

\usepackage{geometry}
\usepackage{amsfonts}
\usepackage{amsmath}
\usepackage{amssymb}
\usepackage[dvips]{color,graphicx}

\newcommand{\be}{\begin{equation}}
\newcommand{\ee}{\end{equation}}

\newcommand{\ba}{\begin{equation}\begin{array}}
\newcommand{\ea}{\end{array}\end{equation}}

\newcommand{\bea}{\begin{eqnarray}}
\newcommand{\eea}{\end{eqnarray}}

\newcommand{\bse}{\begin{subequations}}
\newcommand{\ese}{\end{subequations}}

\newcommand{\Too}{T^{\hat{0}\hat{0}}}
\newcommand{\Trr}{T^{\hat{1}\hat{1}}}
\newcommand{\Tzz}{T^{\hat{2}\hat{2}}}
\newcommand{\Tff}{T^{\hat{3}\hat{3}}}
\newcommand{\Tof}{T^{\hat{0}\hat{3}}}
\newcommand{\Tab}{T^{\hat{a}\hat{b}}}

\newcommand{\rr}{r}

\title{Spinning straight cosmic strings with flat exterior solutions generically violate the weak energy condition}
\author{A J Janca\\\smallskip\footnotesize{Department of Physics, North Carolina State University\\Raleigh NC 27695, United States}}
\date{9 May 2007}

\makeatletter
\renewcommand{\maketitle}{%
  \begin{flushleft}%
    {\huge\bfseries\@title\par}%
    \bigskip \bigskip \bigskip \bigskip \bigskip
    {\large\@author\par}%
    \smallskip
    {\large\@date\par}%
    \bigskip
  \end{flushleft}%
}
\makeatother

\begin{document}
\maketitle

\begin{small}

\noindent Any interior solution for a cylindrically symmetric, stationary cosmic string with flat exterior, spinning around its longitudinal axis, and without internal longitudinal currents ($g_{zz}=1$, $g_{tz}=0$), must somewhere violate the weak energy condition of standard general relativity.  Existing interior solutions may be readily fixed by adding mass to the string above that generating its angular deficit, but at the cost of introducing an exterior gravitational field.\\

\end{small}

\vspace{12pt}

\noindent This note concerns interior solutions for spinning flat-exterior cosmic strings, when modelled as objects of finite thickness, in standard general relativity.  As far as I know, there are only two such examples in the literature \cite{JensenSoleng1992, Krisch1996}; however, the difficulty described below seems to be quite general.\footnote{A good part of \cite{JensenSoleng1992, Krisch1996} and similar papers \cite{Soleng1992, Soleng1994} also discuss interior solutions within an extension to standard general relativity permitting torsion: this note does not cover these extended-theory solutions.}

In the following, where $\rr$ appears as a subscript it represents a partial derivative, $f_{\rr}=\partial f /\partial \rr$, and geometric units are used such that $c$ and $8 \pi G = 1$.\footnote{In the notation of \cite{JensenSoleng1992}, $w=M$, $b=A$, $w_{\rr}/2b = \Omega$, $\rr_1=r_s$, $\rr_c = r_0$, $f_{\rr}=f'$.}

A stationary cosmic string with a flat vacuum exterior can be described by the metric
\vspace{-6pt}
\bea
ds^2 &=& (dt-w \, d\phi)^2 - d\rr^2 - dz^2 - b^2 d\phi^2
\eea
\noindent with $w$ and $b$ both functions of $\rr$. \cite{JensenSoleng1992}  For a string of non-zero thickness to be `regular' (locally Euclidean) on its central axis, its interior solution must have $b(\rr) \to \rr$ and $w \to 0$ as $\rr \to 0$.  For its exterior to be flat, outside some finite coordinate radius $\rr=\rr_1$ marking the boundary of the string $b$ and $w$ must have the form $b=B(\rr+\rr_c)$, $w=w_1$, with $B, \, \rr_c, \, w_1$ constant.  

In the local frame defined by $dx^{\hat{0}}=dt-w\,d\phi$, $dx^{\hat{1}}=d\rr$, $dx^{\hat{2}}=dz$, $dx^{\hat{3}}=b \, d\phi$, the string's energy tensor is
%
%
%
%
\ba{lllll} \label{Tab}
b^2 \Too &=& \!\!\! \phantom{-}\frac{3}{4} \, {w_{\rr}^2} - b\,b_{\rr \rr} &=&  \!\!\! - b^2 \Tzz \\
b^2 \Trr &=&  \!\!\! \phantom{-}\frac{1}{4} \, w_{\rr}^2 &=&  \!\!\! \phantom{-} b^2 \Tff \\
b^2 \Tof &=& \!\!\! -\frac{1}{2} \, b \, w_{\rr \rr} + \frac{1}{2} \, w_{\rr} b_{\rr} 
\ea

\noindent Outside the string, $w_{\rr}, w_{\rr \rr}, b_{\rr \rr}$ vanish and the spacetime is vacuum.  For the static string where $w=0$ everywhere, the energy tensor simplifies to $\Too = - \Tzz = -b_{\rr \rr}/b$, which for $b_{\rr \rr} < 0$ represents a longitudinal tension equal in magnitude to the string's mass density. \cite{Hiscock1985, Gott1985}

\begin{figure}[t]
\begin{centering} \includegraphics[angle=0, width=\columnwidth]{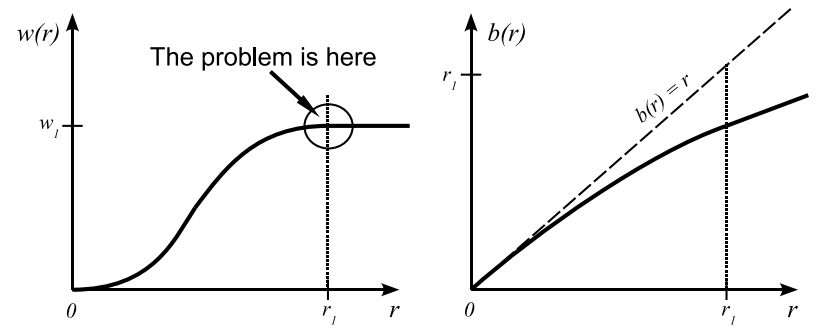}\end{centering}
\caption[General form of $w(\rr)$ and $b(\rr)$]{General form of $w(\rr)$ and $b(\rr)$.  As explained in the text, their specific functional forms aren't relevant.}
\label{fig:wandb}
\end{figure}

For a rotating string, the troublesome boundary condition turns out to be $w_{\rr}=0$ at its external boundary ${\rr}={\rr}_1$, which is necessary to smoothly (at least $C^1$) match $w(\rr)$ to constant $w_1$ outside (figure \ref{fig:wandb}).  A fast timelike observer travelling at constant $\rr=\rr_1$ with (covariant) four-velocity (note that here $\beta$ does \emph{not} denote $v/c$)
%
%
\bea
u_{\hat{a}} &=& \epsilon^{-1} (1, \: 0, \: \alpha, \beta) \\
\alpha^2 + \beta^2 &=& 1 - \epsilon \nonumber
\eea

\noindent will perceive the local energy density of the string to be
\bse \label{wec0string} \bea
\epsilon^2 \Tab u_{\hat{a}} u_{\hat{b}} &=& \Too + \alpha^2 \Tzz + \beta^2 \Tff + 2\beta \Tof \\
&=& b^{-1} [-(\beta^2 + \epsilon) \, b_{\rr \rr} -\beta \, w_{\rr \rr}]
\eea \ese

\noindent For any $\beta$, an observer travelling sufficiently close to $v=c$ can reduce $\epsilon \ll \beta^2$, so that this expression of the weak energy condition (WEC) of classical general relativity may arbitrarily closely approach the null energy condition for a lightlike observer with $\epsilon \rightarrow 0$.\footnote{This trajectory corresponds to that of an observer travelling close to light speed on a path parallel to the $\rr$ axis, and spiralling around it against the rotation of the source.}  By choosing a path with $\beta$ sufficiently small and of the same sign as $w_{\rr \rr}$, (\ref{wec0string}) can be made negative.  Thus any interior solution for a rotating string, varying only the spacetime's angular deficit $b(\rr)/\rr$ and angular velocity $w(\rr)$, must at some point violate the weak energy condition.

Neither the magnitude at $\rr_1$ of $w$ nor its particular functional form within the string affect this result.  In particular, neither enforcing continuity greater than $C^1$ at the boundary (so that $w_{\rr \rr}=0$ at $\rr_1$) nor pulling $\rr_1$ out to spatial infinity will remove the problem (appendix \ref{extensions}).

This violation is easy to miss unless one is looking for it.  For seemingly reasonable values of $w_1 \ll \rr_1$ (well below, say, the threshold necessary to create closed timelike curves) interior solutions can be constructed that will satisfy the WEC for any timelike $u^a$ with spatial component projecting in either one of the ${z}$ or ${\phi}$ directions. \cite{Soleng1994}  The violation becomes apparent only when one considers trajectories with a little bit of both.  

That is not to say that rotating strings can't exist.\footnote{The larger problem of their unrealistic infinite extension in the $\pm z$ direction can be fixed by enclosing them within a suitably bounded object, such as the toroidal trap of \cite{grqc0701085}.}  If there is a difference in $g_{t\phi}$ between the central axis and some other point in the spacetime, the exterior to the string cannot be simply flat.  Some additional field or fields must be added to the metric and energy tensor to compensate for the energy failure close to $\rr_1$, so that $\Too$ will have enough of a mass surplus over the other components to satisfy at least the WEC for all timelike observers (appendix \ref{fixing}).  That part of the mass density of a rotating string due to its angular deficit $\mu = 2\pi (B-1)$ is insufficient by itself to support a physically realistic interior solution.

I am grateful to S V Krasnikov for helpful criticism while writing this note.

\begin{appendix}

\section*{Appendices}
\section{\label{extensions}Adjusting the boundary}

What matters for (\ref{wec0string}) is that either $b\,w_{\rr \rr}$ or $w_{\rr} b_{\rr}$ (if $w \rightarrow 0$ on the central axis $\rr=0$, $w_{\rr}$ and $-w_{\rr \rr}$ must have the same sign at $\rr_1$) must vanish less fast with increasing $\rr$ than $w_{\rr}^2$.  For an interior solution where $w_{\rr \rr}$ also equals 0 at the boundary $\rr_1$, there will still be some point inside the string where $|w_{\rr \rr}| \gg w_{\rr}^2 /b$, and the WEC will be violated there.  And even if the outer boundary of the string were stretched out to $\rr \rightarrow \infty$ (so that the string filled all of space) such that for large $\rr$, $w \sim w_1 - C/\rr^n$, the WEC could not be satisfied for any $n>-1$, which would result in $|w| > C \rr$ as $\rr \rightarrow \infty$\textemdash increasing without bound instead of tapering off. 


\section{Additional fields}\label{fixing}

Adding a small amount of linear mass density $m$ such that the external metric has the stationary Weyl form for an infinite line mass
\be
ds^2 = e^{-2f} (dt-w\,d\phi)^2 - e^{2f} (b^2 d\phi^2 + e^{2k} [d\rr^2+dz^2])
\ee

\noindent with $f= -m \ln (\rr / \rr_c)$ and $k=-m f$, and with some suitably chosen interior $f$ (such that $f$ and $f_{\rr}$ are continuous at $\rr_1$ and $f_{\rr}=0$ at $\rr=0$, and $f_{\rr \rr} \leq 0$ within the string so that $\Too$ is positive and dominant over all combinations of $\Tab$'s other components) would be sufficient to ensure that an interior solution with $b$ and $w$ of the general form here would be physically satisfactory.  Other exact solutions for infinite line sources may be adapted to the spinning string in a similar manner.  An explicit example adapting the van Stockum solution is given in \cite{Krisch2003}. 

\end{appendix}

\end{document}